\documentclass[prd,twocolumn,showpacs,showkeys,floatfix]{revtex4}
\usepackage{amsmath,amssymb}
\usepackage{latexsym}
\usepackage{bm}
\usepackage{hyperref}

\makeatletter
\def\eqnarray{\stepcounter{equation}\let\@currentlabel=\theequation
\global\@eqnswtrue
\global\@eqcnt\z@\tabskip\@centering\let\\=\@eqncr
$$\halign to \displaywidth\bgroup\@eqnsel\hskip\@centering
  $\displaystyle\tabskip\z@{##}$&\global\@eqcnt\@ne
  \hfil${\;##\;}$\hfil
  &\global\@eqcnt\tw@ $\displaystyle\tabskip\z@{##}$\hfil
   \tabskip\@centering&\llap{##}\tabskip\z@\cr}
\makeatother

\begin{document}
\title{CAVITY EVOLUTION IN RELATIVISTIC SELF-GRAVITATING FLUIDS}
\author{L. Herrera}
\email{laherrera@cantv.net.ve}
\affiliation{Escuela de F\'{\i}sica, Facultad de Ciencias,
Universidad Central de Venezuela, Caracas, Venezuela.}
\author{G. Le Denmat}
\email{gerard.le\_denmat@upmc.fr}
\affiliation{Observatoire de Paris, Universit\'e Pierre et Marie Curie,LERMA(ERGA) CNRS - UMR 8112, 94200 Ivry, France.}
\author{N.O. Santos}
\email{N.O.Santos@qmul.ac.uk}
\affiliation{School of Mathematical Sciences, Queen Mary
University of London, London E1 4NS, UK. and\\
Laborat\'orio Nacional de Computa\c{c}\~ao Cient\'{\i}fica,
25651-070 Petr\'opolis RJ, Brazil  and Observatoire de Paris, Universit\'e Pierre et Marie Curie,LERMA(ERGA) CNRS - UMR 8112, 94200 Ivry, France.}

\begin{abstract}
We consider the  evolution of cavities   within spherically symmetric  relativistic fluids, under  the assumption that proper radial distance between  neighboring fluid elements  remains constant during their evolution (purely areal evolution condition). The general formalism is deployed and  solutions are presented. Some of them satisfy Darmois conditions whereas  others present shells and must satisfy Israel conditions, on either one or both boundary surfaces. Prospective applications of these results to some astrophysical scenarios 	is  suggested.

\end{abstract}

\date{\today}
\pacs{04.40.-b, 04.20.-q, 04.40.Dg, 04.40.Nr}
\keywords{Relativistic fluids, voids.}
\maketitle

\newpage

\section{INTRODUCTION}
Many years ago, Skripkin \cite{Skripkin} addressed the very interesting problem of the evolution of a  spherically symmetric fluid  distribution  following a central explosion. As a result of the conditions imposed by Skripkin  a Minkowskian  cavity should surround the centre of the fluid distribution.

Recently \cite{H1}, this problem was studied in detail, proving that under Skripkin conditions  (isotropic fluid with constant energy density distribution) the scalar expansion vanishes. It was  further shown that the assumption of vanishing expansion scalar requires  the existence of a cavity within the fluid distribution (of any kind). Next, it  was shown in \cite{H2}, that the Skripkin model   is incompatible with Darmois junction conditions \cite{26}. Also, the inhomogeneous expansionfree dust  models presented in \cite{H2} are deprived of physical interest since they imply negative energy density distributions.

 For the reasons above, we turn in this paper to another kinematical condition particularly suitable for describing the evolution of a  fluid distribution with a cavity surrounding the centre. This consists in assuming  the vanishing  of variation of proper radial distance between any two infinitesimally close fluid elements per unit of proper time.  We shall explore here the consequences derived from this condition (hereafter referred to as the purely areal evolution condition).  In particular we are interested in analytical models which even if are relatively simple to analyze, still contain some of the essential features of a realistic situation.  It should be emphasized that we are not interested in the dynamics and the conditions of the creation of the cavity itself, but only in its evolution once  it is already formed.

We have two hypersurfaces delimiting the fluid. The external one separating the fluid distribution from a Schwarzschild or Vaidya spacetime (depending on whether we assume the evolution to be adiabatic or dissipative) and the  internal one, delimiting the cavity within which we have Minkowski spacetime. It should be mentioned that for cavities with sizes of the order of 20 Mpc or smaller, the assumption of  a  spherically symmetric spacetime outside the cavity is quite reasonable, since the observed universe cannot be considered homogeneous on scales less than 150-300 Mpc. However for  larger cavities it should be more appropriate to consider their embedding in an expanding Lema\^{\i}tre-Friedmann-Robertson-Walker spacetime (for the specific case of void modeling in expanding universes see \cite{bill}, \cite{Torres} and references therein).

Thus,  we have to consider junction conditions on both hypersurfaces. Depending on whether we impose Darmois conditions \cite{26} or allow for the existence of thin shells \cite{Israel}, different kind of models are obtained. In this paper we shall focus mainly on models satisfying Darmois conditions, although some models presenting thin shells will be briefly  described too.

For sake of generality we shall start our discussion by  considering an anisotropic  dissipative fluid (arguments to justify such kind of fluid distributions may found in \cite{Herreraanis}-\cite{Hetal}  and references therein). A detailed description of  this kind of distribution, as well as definitions of kinematical and another important variables, are given in section II.   The Darmois junction conditions on both, the inner and the outer boundary surface, are briefly discussed in section III.

In order to understand better the physical meaning of the purely areal evolution condition, we shall discuss two different definitions of  radial  velocity of a fluid element, in terms of which both the expansion and the shear can be expressed, and which renders intelligible the origin of the term we use to denote such a  condition in section IV.

We shall next deploy all the equations required for cavity modeling under the purely areal evolution condition in section V. The specific case of  cavities  satisfying Darmois conditions, on both hypersurfaces, is treated in  section VI, whereas cavities presenting shells are considered in section VII.

Finally a brief summary of the results is presented in the last section and prospective applications of these results are briefly mentioned.

\section{FLUID DISTRIBUTIONS AND KINEMATICAL VARIABLES}
We consider a spherically symmetric distribution  of
fluid, bounded by a spherical surface $\Sigma^{(e)}$. The fluid is
assumed to be locally anisotropic, with principal stresses unequal, and undergoing dissipation in the
form of heat flow (diffusion approximation).

Choosing comoving coordinates inside $\Sigma^{(e)}$, the general
interior metric can be written
\begin{equation}
ds^2_-=-A^2dt^2+B^2dr^2+R^2(d\theta^2+\sin^2\theta d\phi^2),
\label{1}
\end{equation}
where $A$, $B$ and $R$ are functions of $t$ and $r$ and are assumed
positive. We number the coordinates $x^0=t$, $x^1=r$, $x^2=\theta$
and $x^3=\phi$. Observe that $A$ and $B$ are dimensionless, whereas $R$ has the same dimension as $r$.

From (\ref{1}) we have that inside $\Sigma^{(e)}$ any spherical surface has its proper radius given by $\int Bdr$ and its areal radius by $R$.

The matter energy-momentum $T_{\alpha\beta}^-$ inside $\Sigma^{(e)}$
has the form
\begin{equation}
T_{\alpha\beta}^-=(\mu +
P_{\perp})V_{\alpha}V_{\beta}+P_{\perp}g_{\alpha\beta}+(P_r-P_{\perp})\chi_{
\alpha}\chi_{\beta}+q_{\alpha}V_{\beta}+V_{\alpha}q_{\beta}, \label{2}
\end{equation}
where $\mu$ is the energy density, $P_r$ the radial pressure,
$P_{\perp}$ the tangential pressure, $q^{\alpha}$ the heat flux,
$V^{\alpha}$ the four-velocity of the fluid and
$\chi^{\alpha}$ a unit four-vector along the radial direction.
These quantities satisfy
\begin{equation}
V^{\alpha}V_{\alpha}=-1, \;\; V^{\alpha}q_{\alpha}=0, \;\; \chi^{\alpha}\chi_{\alpha}=1, \;\;
\chi^{\alpha}V_{\alpha}=0. \label{3}
\end{equation}
The four-acceleration $a_{\alpha}$ and the expansion $\Theta$ of the fluid are
given by
\begin{equation}
a_{\alpha}=V_{\alpha ;\beta}V^{\beta}, \;\;
\Theta={V^{\alpha}}_{;\alpha}, \label{4}
\end{equation}
and its  shear $\sigma_{\alpha\beta}$ by
\begin{equation}
\sigma_{\alpha\beta}=V_{(\alpha
;\beta)}+a_{(\alpha}V_{\beta)}-\frac{1}{3}\Theta h_{\alpha \beta},
\label{5}
\end{equation}
where
\begin{equation}
h_{\alpha \beta}=g_{\alpha\beta}+V_{\alpha}V_{\beta}.
\label{proy}
\end{equation}

Since we assumed the metric (\ref{1}) comoving then
\begin{equation}
V^{\alpha}=A^{-1}\delta_0^{\alpha}, \;\;
q^{\alpha}=qB^{-1}\delta^{\alpha}_1, \;\;
\chi^{\alpha}=B^{-1}\delta^{\alpha}_1, \label{6}
\end{equation}
where $q$ is a function of $t$ and $r$.
From  (\ref{4}) with (\ref{6}) we have the non zero component for the  four-acceleration and its scalar,
\begin{equation}
a_1=\frac{A^{\prime}}{A}, \;\; a=(a^{\alpha}a_{\alpha})^{1/2}=\frac{A^{\prime}}{AB}, \label{7}
\end{equation}
and for the expansion
\begin{equation}
\Theta=\frac{1}{A}\left(\frac{\dot{B}}{B}+2\frac{\dot{R}}{R}\right),
\label{8}
\end{equation}
where the  prime stands for $r$
differentiation and the dot stands for differentiation with respect to $t$.
With (\ref{6}) we obtain
for the shear (\ref{5}) its non zero components
\begin{equation}
\sigma_{11}=\frac{2}{3}B^2\sigma, \;\;
\sigma_{22}=\sigma_{33}\sin^{-2}\theta=-\frac{1}{3}R^2\sigma,
 \label{10}
\end{equation}
and its scalar
\begin{equation}
\sigma^{\alpha\beta}\sigma_{\alpha\beta}=\frac{2}{3}\sigma^2,
\label{5bn}
\end{equation}
where
\begin{equation}
\sigma=\frac{1}{A}\left(\frac{\dot{B}}{B}-\frac{\dot{R}}{R}\right).\label{11}
\end{equation}

Also observe that the shear tensor may be written as:
\begin{equation}
\sigma_{\alpha \beta}= \sigma \left(\chi_\alpha \chi_\beta - \frac{1}{3} h_{\alpha \beta}\right).
\label{sh}
\end{equation}

Sometimes it could be convenient \cite{Mar, Hshe} to express the energy--momentum tensor  (\ref{2}) in the form
\begin{equation}
T_{\alpha \beta}^- = \mu V_\alpha V_\beta + \hat{P} h_{\alpha \beta} + \Pi_{\alpha \beta} +
q \left(V_\alpha \chi_\beta + \chi_\alpha V_\beta\right)
\label{Tabs}
\end{equation}
with
$$\hat P=\frac{1}{3}h_{\alpha \beta} T^{\alpha \beta}=\frac{P_{r}+2P_{\bot}}{3},$$
$$\Pi^{\alpha \beta}=\left(h^{(\alpha}_\gamma h^{\beta)}_\delta-\frac{1}{3}h^{\alpha \beta}h_{\gamma \delta}\right)T^{\gamma \delta}=\Pi\left(\chi^\alpha \chi^\beta - \frac{1}{3} h^{\alpha \beta}\right),$$
$$\Pi=P_{r}-P_{\bot}.$$

Next, the mass function $m(t,r)$ introduced by Misner and Sharp
\cite{Misner} (see also \cite{Cahill}) reads
\begin{equation}
m=\frac{R^3}{2}{R_{23}}^{23}
=\frac{R}{2}\left[\left(\frac{\dot R}{A}\right)^2-\left(\frac{R^{\prime}}{B}\right)^2+1\right].
 \label{18}
 \end{equation}
To study the dynamical properties of the system, let us  introduce,
following Misner and Sharp \cite{Misner}, the proper time derivative $D_T$
given by
\begin{equation}
D_T=\frac{1}{A}\frac{\partial}{\partial t}. \label{16N}
\end{equation}

Using (\ref{16N}) we can define the velocity $U$ of the collapsing
fluid (for another definition of velocity see  section IV) as the variation of the areal radius with respect to proper time, i.e.,
\begin{equation}
U=D_TR<0 \;\; \mbox{(in the case of collapse)}. \label{19}
\end{equation}
Then (\ref{18}), by using (\ref{19}), can be rewritten as
\begin{equation}
E \equiv \frac{R^{\prime}}{B}=\left(1+U^2-\frac{2m}{R}\right)^{1/2}.
\label{20x}
\end{equation}

Using  field equations (see \cite{H1} for details) ) with (\ref{16N}) and (\ref{19}) we obtain from
(\ref{18})
\begin{eqnarray}
m^{\prime}=4\pi\left(\mu+q \frac{U}{E}\right)R^{\prime}R^2,
\label{27Dr}
\end{eqnarray}
which implies
\begin{equation}
m=4\pi\int^{r}_{0}\left(\mu +q\frac{U}{E}\right)R^2R^{\prime}dr, \label{27intcopy}
\end{equation}
where we assumed a regular centre to the distribution, so $m(0)=0$.

It will be useful to introduce the  Weyl tensor. Thus,
let  $E_{\alpha\beta}$ denote the ``electric'' part of the Weyl tensor (in the spherically symmetric case the ``magnetic'' part of the Weyl tensor
vanishes, $H_{\alpha \beta}=0$) defined by
\begin{equation}
E_{\alpha \beta} = C_{\alpha \mu \beta \nu} V^\mu V^\nu,
\label{elec}
\end{equation}
which may be written as
\begin{equation}
E_{\alpha \beta}={\cal E}\left(\chi_\alpha \chi_\beta-\frac{1}{3}h_{\alpha \beta}\right),
\label{52}
\end{equation}
where
\begin{widetext}
\begin{eqnarray}
{\cal E}= \frac{1}{2 A^2}\left[\frac{\ddot R}{R} - \frac{\ddot B}{B} - \left(\frac{\dot R}{R} - \frac{\dot B}{B}\right)\left(\frac{\dot A}{A} + \frac{\dot R}{R}\right)\right]
+\frac{1}{2 B^2} \left[\frac{A^{\prime\prime}}{A} - \frac{R^{\prime\prime}}{R} + \left(\frac{B^{\prime}}{B} + \frac{R^{\prime}}{R}\right)\left(\frac{R^{\prime}}{R}-\frac{A^{\prime}}{A}\right)\right]  -\frac{1}{2 R^2}.
\label{E}
\end{eqnarray}
\end{widetext}

Using field equations (see \cite{H1} for details) and the definition of mass function (\ref{18})  we may write $\cal E$ as
\begin{equation}
{\cal E}=4\pi (\mu-P_r+P_{\perp})-\frac{3m}{R^3}.
\label{W1}
\end{equation}

\section{THE EXTERIOR SPACETIME AND JUNCTION CONDITIONS}
Outside $\Sigma^{(e)}$ we assume we have the Vaidya
spacetime (or Schwarzschild in the dissipationless case), i.e., we assume all outgoing radiation is massless,
described by
\begin{equation}
ds^2=-\left[1-\frac{2M(v)}{r}\right]dv^2-2drdv+r^2(d\theta^2
+\sin^2\theta
d\phi^2) \label{19d},
\end{equation}
where $M(v)$  denotes the total mass,
and  $v$ is the retarded time.
The matching of the non-adiabatic sphere to
the Vaidya spacetime, on the surface $r=r_{\Sigma^{(e)}}=$ constant, in the absence of thin shells, where Darmois conditions hold, is discussed in \cite{matter, Santos, Bonnor, chan1}. This requires the continuity of the first and the second fundamental forms through the matching hypersurface, producing
\begin{equation}
m(t,r)\stackrel{\Sigma^{(e)}}{=}M(v), \label{20}
\end{equation}
\begin{widetext}
\begin{eqnarray}
2\left(\frac{{\dot R}^{\prime}}{R}-\frac{\dot B}{B}\frac{R^{\prime}}{R}-\frac{\dot R}{R}\frac{A^{\prime}}{A}\right)
\stackrel{\Sigma^{(e)}}{=} -\frac{B}{A}\left[2\frac{\ddot R}{R}
-\left(2\frac{\dot A}{A}
-\frac{\dot R}{R}\right)\frac{\dot R}{R}\right]+\frac{A}{B}\left[\left(2\frac{A^{\prime}}{A}
+\frac{R^{\prime}}{R}\right)\frac{R^{\prime}}{R}-\left(\frac{B}{R}\right)^2\right],
\label{21}
\end{eqnarray}
\end{widetext}
and
\begin{equation}
q\stackrel{\Sigma^{(e)}}{=}\frac{L}{4\pi r}, \label{20lum}
\end{equation}
where $\stackrel{\Sigma^{(e)}}{=}$ means that both sides of the equation
are evaluated on $\Sigma^{(e)}$ and $L$ denotes   the total luminosity of the  sphere as measured on its surface and is given by
\begin{equation}
L=L_{\infty}\left(1-\frac{2m}{r}+2\frac{dr}{dv}\right)^{-1}, \label{14a}
\end{equation}
and where
\begin{equation}
L_{\infty}=\frac{dM}{dv} \label{14b}
\end{equation}
is the total luminosity measured by an observer at rest at infinity.

From  (\ref{21}) and field equations  one obtains
\begin{equation}
q\stackrel{\Sigma^{(e)}}{=}P_r.\label{22}
\end{equation}

In the case when a cavity forms, then we also have to match the solution to the Minkowsky spacetime on the boundary surface delimiting the cavity. If we call $\Sigma^{(i)} $ the boundary surface between the cavity and the fluid, then the matching of the Minkowski spacetime within the cavity to the fluid distribution, implies
\begin{equation}
m(t,r)\stackrel{\Sigma^{(i)}}{=}0, \label{junction1i}
\end{equation}
\begin{equation}
q\stackrel{\Sigma^{(i)}}{=}P_r.\label{j3i}
\end{equation}
However, since we are assuming our  cavity to be empty, then $L\stackrel{\Sigma^{(i)}}{=}0$, which implies
\begin{equation}
q\stackrel{\Sigma^{(i)}}{=}P_r\stackrel{\Sigma^{(i)}}{=}0.\label{j3in}
\end{equation}

If we allow for the presence of thin shells on $\Sigma^{(i)}$ and/or $\Sigma^{(e)}$, then we have to relax the above conditions and allow for discontinuities in the mass function \cite{Israel}.

\section{TWO DEFINITIONS OF RADIAL VELOCITY AND THE PURELY AREAL EVOLUTION CONDITION}
In section II we introduced the  variable $U$ which, as mentioned before, measures the variation of the areal radius $R$  per unit proper time.

Another possible definition of ``velocity'' may be introduced, as    the variation of the infinitesimal proper radial distance between two neighboring points ($\delta l$) per unit of proper time, i.e. $D_T(\delta l)$.
Then, it can be shown that (see \cite{H1} for details)
\begin{equation}
 \frac{D_T(\delta l)}{\delta l}= \frac{1}{3}(2\sigma +\Theta),
\label{vel15}
\end{equation}
or, by using (\ref{8}) and (\ref{11}),
\begin{equation}
 \frac{D_T(\delta l)}{\delta l}= \frac{\dot B}{AB}.
\label{vel16}
\end{equation}
Then with (\ref{8}), (\ref{11}), (\ref{19}) and (\ref{vel16}) we can write
\begin{equation}
 \sigma= \frac{D_T(\delta l)}{\delta l}-\frac{D_T R}{R}= \frac{D_T(\delta l)}{\delta l}-\frac{U}{R},
\label{vel17}
\end{equation}
and
\begin{equation}
 \Theta= \frac{D_T(\delta l)}{\delta l}+\frac{2D_T R}{R}=\frac{D_T(\delta l)}{\delta l}+\frac{2U}{R},
\label{vel17bis}
\end{equation}
Thus the ``circumferential'' (or ``areal'') velocity $U$,  is  related to the change  of areal radius $R$  of a layer of matter, whereas $D_T(\delta l)$, has also the meaning of ``velocity'', being the relative velocity between neighboring layers of matter, and can be in general different from $U$.

In \cite{H1} it was shown that the condition $\Theta=0$ requires the existence of a cavity surrounding the centre of the fluid distribution.

Let us now consider the condition  $D_T(\delta l)=0$, but $U\neq 0$. From the  comments above it is evident why we shall refer to it as  the purely areal evolution condition.

Now,   if $D_T(\delta l)=0$ then  it follows from (\ref{vel17}) and (\ref{vel17bis}) that $\Theta=-2\sigma$, feeding this  back into
the $(01)$ component of the Einstein field equations (see \cite{H1}) for details)  we get
\begin{equation}
\sigma^{\prime}+\frac{\sigma R^{\prime}}{R}=-\frac{4\pi q R^{\prime}}{E},
\label{27}
\end{equation}
whose integration with respect to $r$ yields
\begin{equation}
\sigma=\frac{\zeta(t)}{R}-\frac{4\pi}{R}\int^{r}_{0} q\frac{RR^{\prime}}{E}\;dr,
\label{28relvel}
\end{equation}
where $\zeta$ is an integration function of  $t$. It should be observed that in the case where the fluid fill all the sphere, including the centre ($r=0$), we should impose the regularity condition $\zeta=0$. However since we consider the possibility of  a cavity surrounding the centre, such a condition is not required.

On the other hand, (\ref{28relvel}) with (\ref{vel17}) implies
\begin{equation}
U=-\zeta+4\pi\int^{r}_{0} q\frac{RR^{\prime}}{E}\;dr.
\label{29relvel}
\end{equation}
Thus, in the non-dissipative case the purely areal evolution condition implies that $U=U(t)$. This condition is clearly incompatible with  a regular symmetry centre, unless $U= 0$.Therefore if we want the purely areal evolution condition to be compatible with a time dependent situation ($U\neq0$) we must assume that  either 
\begin{itemize}
\item the fluid has no symmetry centre,\\
or
\item the centre is surrounded by  a compact spherical section of another spacetime, suitably matched to the rest of the fluid.
\end{itemize}

Here we shall discard the first possibility since we are particularly interested in describing localized objects without the unusual topology of a spherical fluid without a center. Also, within the context of the second alternative we have chosen an inner vacuum Minkowski spherical vacuole.

Let us now consider  the dissipative case. Then assuming   the purely areal evolution condition,  if the fluid fills the whole sphere (no cavity surrounding the centre),  and we have a symmetry centre,  we have to put $\zeta=0$ and (\ref{29relvel}) becomes
\begin{equation}
U=4\pi\int^{r}_{0} q\frac{RR^{\prime}}{E}\;dr,
\label{29relvelbis}
\end{equation}
which is not incompatible with a regular symmetry centre. In this case we shall assume in an {\it ad hoc} manner that a cavity surrounds the centre. However, this assumption is somehow suggested by the following qualitative argument. 

In the case of an outwardly directed flux vector ($q>0$), all terms within the integral are positive and  we obtain from (\ref{vel17bis}) and (\ref{29relvelbis}) that $\Theta>0$ and $U>0$. Now, during the Kelvin-Helmholtz phase of evolution  \cite{KW}, when all the dissipated energy comes from the gravitational energy, we should expect  a  contraction, not expansion, to be associated with an outgoing dissipative flux. Inversely,  an inwardly directed flux ($q<0$)  (during that phase) would produce an overall expansion, not a contraction as it follows from (\ref{29relvelbis}).

Thus we have seen that the purely areal evolution condition appears to be  particularly suitable to describe  the evolution of a fluid distribution with  a cavity surrounding the centre.

Finally observe that  using (\ref{vel17}) and (\ref{vel17bis}) in (\ref{sh}), the purely areal evolution condition can be expressed in a covariant form as:
\begin{equation}
\sigma_{\alpha \beta}= -\frac{\Theta}{2} \left(\chi_\alpha \chi_\beta - \frac{1}{3} h_{\alpha \beta}\right).
\label{shbis}
\end{equation}

In the next section we shall consider some models.

\section{MODELS OF CAVITIES}
We shall now study the general properties of  models satisfying the purely areal evolution condition.

The general picture is similar to that proposed by Skripkin \cite{Skripkin}, namely, an explosion at the center initiates an overall expansion of the fluid, creating a cavity surrounding the centre. The difference here is that we shall not assume $\Theta=0$ but instead, $D_T(\delta l)=0$.
Thus we have because of (\ref{vel16}) ${\dot B}=0$ (but  ${\dot R}\neq 0$ ) which means that $B=B(r)$ and it can be chosen
\begin{equation}
B=1, \label{23}
\end{equation}
with no loss of generality.  As mentioned before, the  physical appeal of this kind of models stems from the fact that the condition ${\dot B}=0$  requires for consistency, the existence of a cavity surrounding the centre.

Then the Einstein field equations  become
\begin{eqnarray}
8\pi\mu=\frac{1}{A^2}\left(\frac{\dot R}{R}\right)^2
-2\frac{R^{\prime\prime}}{R}-\left(\frac{R^{\prime}}{R}\right)^2+\frac{1}{R^2},\label{24a}
\end{eqnarray}
\begin{eqnarray}
8\pi q=\frac{2}{A}\left(\frac{{\dot R}^{\prime}}{R}-\frac{\dot R}{R}\frac{A^{\prime}}{A}\right), \label{25a}
\end{eqnarray}
\begin{eqnarray}
8\pi P_r=-\frac{1}{A^2}\left[2\frac{\ddot R}{R}-\left(2\frac{\dot A}{A}
-\frac{\dot R}{R}\right)\frac{\dot R}{R}\right]\nonumber \\+\left(2\frac{A^{\prime}}{A}
+\frac{R^{\prime}}{R}\right)\frac{R^{\prime}}{R}  -\frac{1}{R^2}, \label{26a}
\end{eqnarray}
\begin{eqnarray}
8\pi P_{\perp}=-\frac{1}{A^2}\left(\frac{\ddot R}{R}
-\frac{\dot A}{A}\frac{\dot R}{R}\right)+\frac{A^{\prime\prime}}{A}+\frac{R^{\prime\prime}}{R}
+\frac{A^{\prime}}{A}\frac{R^{\prime}}{R}, \label{27a}
\end{eqnarray}
and the non trivial components of the Bianchi identities, $T_{;\beta}^{-\alpha\beta}=0$, become,
\begin{eqnarray}
\frac{1}{A}\left[{\dot \mu}+2(\mu +P_{\perp})\frac{\dot R}{R}\right]+q^{\prime}+2q\frac{(AR)^{\prime}}{AR}=0, \label{28n}
\end{eqnarray}
\begin{eqnarray}
\frac{1}{A}\left({\dot q}+2q\frac{\dot R}{R}\right)+P_r^{\prime}+(\mu +P_r)\frac{A^{\prime}}{A}\nonumber \\+2(P_r-P_{\perp})\frac{R^{\prime}}{R}=0. \label{29}
\end{eqnarray}

In the particular  geodesic case  we have $A^\prime=0 \rightarrow A=1$. Then the  field equations (\ref{24a}-\ref{27a})  read
\begin{eqnarray}
8\pi\mu=\left(\frac{\dot R}{R}\right)^2
-2\frac{R^{\prime\prime}}{R}-\left(\frac{R^{\prime}}{R}\right)^2+\frac{1}{R^2},\label{24}\\
8\pi q=2\frac{{\dot R}^{\prime}}{R}, \label{25}\\
8 \pi P_r=-\left[2\frac{\ddot R}{R}
+\left(\frac{\dot R}{R}\right)^2\right]
+\left(\frac{R^{\prime}}{R}\right)^2-\frac{1}{R^2}, \label{26}\\
8\pi P_{\perp}=-\frac{\ddot R}{R}+\frac{R^{\prime\prime}}{R}, \label{27bis}
\end{eqnarray}
and it follows with (\ref{18})
\begin{equation}
2\pi \mu=\frac{m}{R^3}+2\pi (P_r-2 P_{\perp}),
\label{30bis}
\end{equation}
or, using (\ref{W1}),
\begin{equation}
2\pi \mu=2\pi (P_r-4P_{\perp})-{\cal E}.
\label{31bis}
\end{equation}
Hence, from (\ref{30bis}) and (\ref{31bis}) we have for conformally flat spacetime
${\cal E}=0$, and geodesic fluids with isotropic pressures $P_r=P_\bot=P$,
\begin{equation}
\frac{m}{R^3}+4\pi P=0,
\label{30bbis}
\end{equation}
implying that such models, if they satisfy Darmois conditions, must be dissipative (otherwise $M=0$), and  absorbing energy ($q<0$, otherwise $m<0$).

\section{Models satisfying Darmois conditions}
Let us now consider some simple cases in order to find  analytical models  which do not present thin shells on either $\Sigma^{(e)}$ or $\Sigma^{(i)}$, but holding Darmois conditions.

The simplest models of this kind we have found are non-dissipative.
Thus, considering $q=0$ then from (\ref{25a}) after integration we have
\begin{equation}
A=\frac{\dot R}{h_1}, \label{n1}
\end{equation}
where $h_1(t)$ is an arbitrary function of $t$.  Reparametrizing $t$, we may choose without loss of generality
\begin{equation}
h_1=\dot R_{\Sigma^{(i)}},
\label{new1a}
\end{equation}
which amounts to choose
\begin{equation}
A_{\Sigma^{(i)}}=1.
\label{new2a}
\end{equation}

Observe that for all these models the velocity $U$ is the same for all fluid elements, since, as it follows from (\ref{19}) and (\ref{n1})
\begin{equation}
U=h_1=\dot R_{\Sigma^{(i)}}. \label{n1a}
\end{equation}
This fact was already brought out in the previous section from (\ref{29relvel}).

Substituting (\ref{n1}) into (\ref{24a}), ({\ref{26a}) and (\ref{27a}) we obtain, using (\ref{16N}), (\ref{n1}) and (\ref{new1a})
\begin{eqnarray}
8\pi\mu=-\frac{1}{R^2}(2RR^{\prime\prime}+R^{\prime 2}-\dot R_{\Sigma^{(i)}}^2-1), \label{n2}\\
8\pi P_r=\frac{1}{R^2{\dot R_{\Sigma^{(i)}}}}D_T[R(R^{\prime 2}-\dot R_{\Sigma^{(i)}}^2-1)], \label{n3}\\
8\pi P_{\perp}=\frac{1}{2R\dot R_{\Sigma^{(i)}}}D_T(2RR^{\prime\prime}+R^{\prime 2}-\dot R_{\Sigma^{(i)}}^2-1). \label{n4}
\end{eqnarray}
From (\ref{n2}) and (\ref{n4}) it is clear that
\begin{equation}
P_{\perp}=-\frac{D_T( \mu R^2)}{2R\dot R_{\Sigma^{(i)}}}. \label{n5}
\end{equation}

Calculating the mass function (\ref{18}) with (\ref{23}) and (\ref{n1}) it yields
\begin{equation}
m=-\frac{R}{2}(R^{\prime 2}-\dot R_{\Sigma^{(i)}}^2-1), \label{n6}
\end{equation}
which allows to reexpress (\ref{n3}) like
\begin{equation}
4\pi P_r=-\frac{\dot m}{R^2\dot R}. \label{n7}
\end{equation}
The boundary conditions $P_r\stackrel{\Sigma^{(e)}}{=}0$ and $P_r\stackrel{\Sigma^{(i)}}{=}0$ are automatically satisfied with $m\stackrel{\Sigma^{(e)}}{=}M=$ constant and $m\stackrel{\Sigma^{(i)}}{=}0$.

With (\ref{n1}) we obtain for (\ref{E})
\begin{equation}
{\cal E}=\frac{R}{4\dot R_{\Sigma^{(i)}}}D_T\left[\frac{1}{R^2}(2RR^{\prime\prime}-R^{\prime 2}+\dot R_{\Sigma^{(i)}}^2+1)\right]. \label{n8}
\end{equation}

We shall next specialize to some particular cases.

\subsection{Conformally flat models}
If we assume the spacetime between $r=r_{\Sigma^{(e)}}$ and $r=r_{\Sigma^{(i)}}$ to be conformally flat ${\mathcal E}=0$, then it follows from (\ref{n8}) that
\begin{equation}
2RR^{\prime\prime}+R^{\prime 2}- f_1 R^2+\dot R_{\Sigma^{(i)}}^2+1=0, \label{ncc10}
\end{equation}
where $f_1(r)$ is an arbitrary function of $r$, and integrating  again we obtain
\begin{equation}
R^{\prime 2}=R\left(\int f_1dR+h_2\right)+\dot R_{\Sigma^{(i)}}^2+1, \label{ncc16}
\end{equation}
where $h_2(t)$ is an arbitrary function of $t$.
Comparing (\ref{n6}) with (\ref{ncc16}) it follows that
\begin{equation}
m=-\frac{R^2}{2}\left(\int f_1 dR +h_2\right), \label{ncc6}
\end{equation}
therefore $h_2$ may be obtained from the junction condition (\ref{junction1i}), producing
\begin{equation}
h_2\stackrel{\Sigma^{(i)}}{=}-\int f_1 dR . \label{ncc7}
\end{equation}
Thus all models of this kind are defined by a single function $f_1(r)$ which should be chosen such as to satisfy  the remaining Darmois conditions.

Before proceeding further, the following remark is in order: all spherically symmetric conformally flat spacetimes (without dissipation) and isotropic fluids are shear-free (see eq.(78) in \cite{Hetal}), but this is no longer true for anisotropic fluids (see eq.(84) in \cite{Hetal}). Therefore the models to be considered here are necessarily anisotropic.

\subsection{Models with vanishing tangential stresses}
Assuming $P_{\perp}=0$ we have from (\ref{n5}) after integration
\begin{equation}
\mu=\frac{f_2}{R^2}, \label{n9}
\end{equation}
where $f_2(r)$ is an arbitrary function of $r$. Substituting (\ref{n9}) into (\ref{n2})  it yields
\begin{equation}
2RR^{\prime\prime}+R^{\prime 2}+8\pi f_2-\dot R_{\Sigma^{(i)}}^2-1=0, \label{n10}
\end{equation}
and with (\ref{n6}) it becomes
\begin{equation}
m^{\prime}=4\pi f_2R^{\prime}. \label{n11}
\end{equation}

In order to obtain models we have to assume specific form of the mass function or the energy density distribution. As an example let us assume
\begin{equation}
f_2=c_1 =\mbox{constant}>0, \label{n12}
\end{equation}
 producing because (\ref{27intcopy}) and (\ref{n9})
\begin{equation}
m=4\pi c_1(R-R_{\Sigma^{(i)}}), \label{n13}
\end{equation}
and implying
\begin{eqnarray}
M=4\pi c_1(R_{\Sigma^{(e)}}-R_{\Sigma^{(i)}}), \label{n14}\\
{\dot R}_{\Sigma^{(e)}}={\dot R}_{\Sigma^{(i)}}, \;\; A_{\Sigma^{(i)}}=A_{\Sigma^{(e)}}=1. \label{n15}
\end{eqnarray}

Also, from (\ref{n7}), (\ref{n9}) and (\ref{n13})  it follows
\begin{equation}
 P_r=\mu(\frac{{\dot R}_{\Sigma^{(i)}} }{\dot R}-1). \label{n7bis}
\end{equation}
Next, substituting (\ref{n13}) into (\ref{n6}) we have
\begin{equation}
RR^{\prime 2}=\alpha R+\beta, \label{n16}
\end{equation}
where
\begin{equation}
\alpha(t)=\dot R_{\Sigma^{(i)}}^2+1-8\pi c_1, \;\; \beta(t)=8\pi c_1 R_{\Sigma^{(i)}}, \label{n17}
\end{equation}
and after integration
\begin{equation}
[\alpha R(\alpha R+\beta)]^{1/2}-\beta\ln[(\alpha R)^{1/2}+(\alpha R+\beta)^{1/2}]=\alpha^{3/2}[r-r_0], \label{n18}
\end{equation}
where $r_0(t)$ is an arbitrary function of $t$.
Evaluating   (\ref{n18}) on $\Sigma^{(i)}$    we obtain
\begin{widetext}
\begin{eqnarray}
\left\{\left[\alpha(\dot R_{\Sigma^{(i)}}^2+1)\right]^{1/2}-8\pi c_1\ln\left[\alpha^{1/2}+(\dot R_{\Sigma^{(i)}}^2+1)^{1/2}\right] -4\pi c_1\ln R\right\}R
\stackrel{\Sigma^{(i)}}{=}a^{3/2}(r-r_0). \label{n19}
\end{eqnarray}
\end{widetext}
This is a first order differential equation for $R_{\Sigma^{(i)}}$ which may be solved for any function $r_0(t)$. The result of this integration, together with (\ref{n18}) provide all the information required to obtain the $t$ and $r$ dependence of all physical and metric variables. Observe that the energy density is always positive and regular everywhere within the fluid distribution. Also, choosing  $r_0(t)$ such that  $0<\frac{\dot R_{\Sigma^{(i)}}}{\dot R}-1\leq1$ we assure that the pressure is positive and smaller that the energy density.

\section{Models with thin shells}
As it  is apparent from the discussion above, the fulfilment  of Darmois conditions on, both, $\Sigma^{(e)}$ and  $\Sigma^{(i)}$ severely  restrict the  possible models of cavities. Therefore it  might be  pertinent to relax Darmois conditions and work within the thin wall  approximation, which allows for the existence of  discontinuities of the mass function across $\Sigma^{(e)}$ and/or  $\Sigma^{(i)}$ (for models of voids within the thin wall approximation see  \cite{v4}-\cite{v6} and references therein).

\subsection{Non-dissipative geodesic model}
The simplest model of this kind (under the purely areal evolution condition) corresponds to a  Lema\^{\i}tre-Tolman-Bondi (LTB)  spacetime \cite{L}-\cite{Bo}, whose general line element is given by
\begin{equation}
ds^2_-=-A^2dt^2+\frac{R^{\prime 2}}{1-K}dr^2+R^2(d\theta^2+\sin^2\theta d\phi^2),
\label{1b}
\end{equation}
where K is a function of $r$.
Then, the purely areal evolution condition applied to (\ref{1b}) produces
\begin{equation}
R^{\prime}=(1-K)^{1/2},
\label{ltb1}
\end{equation}
and since  all LTB spacetimes are geodesic and dissipationless,  (\ref{25}) implies
\begin{equation}
R=h_3+f_3,
\label{32bis}
\end{equation}
where $h_3(t)$ and $f_3(r)$ are arbitrary functions of $t$ and $r$ respectively.
Then for the mass function (\ref{18}) we obtain
\begin{equation}
m=\frac{R}{2}(\dot h_3^2 -f_3^{{\prime}2}+1).
\label{33}
\end{equation}

Imposing Darmois conditions on $\Sigma^{(i)}$ we obtain from (\ref{junction1i}) with (\ref{ltb1}-\ref{33}),
\begin{equation}
h_3=(-K_{\Sigma^{(i)}})^{1/2} t+c_2,
\label{34bis}
\end{equation}
where $c_2$ is an  arbitrary constant and  we have to assume the condition $K_{\Sigma{^(i)}}< 0$.
Once we have imposed Darmois conditions  on $\Sigma^{(i)}$, it follows that we have to assume the presence of a  shell on $\Sigma^{(e)}$.
Indeed, the continuity of the mass function $m$ on $\Sigma^{(e)}$ implies  from  (\ref{20})
\begin{equation}
R\stackrel{\Sigma^{(e)}}{=}\frac{2M}{\dot h_3^2 -f_3^{{\prime}2}+1},
\label{39}
\end{equation}
which, because of (\ref{34bis}) and  the fact that $M$ is constant, produces
\begin{equation}
\dot R\stackrel{\Sigma^{(e)}}{=}0,
\label{35bis}
\end{equation}
implying, because of (\ref{32bis}),
\begin{equation}
\dot h_3=0\rightarrow \dot R=0,
\label{36bis}
\end{equation}
thereby invalidating the continuity of mass function across $\Sigma^{(e)}$. Thus all models of this kind should admit a thin shell on $\Sigma^{(e)}$.
Alternatively, we may assume  discontinuities of the mass function on $\Sigma^{(i)}$ or on both boundaries.

\subsection{Geodesic radiating dust models}
In an increasing order of complexity,  let  us consider the next simplest possible situation, namely, geodesic $a=0$, dust $P_r=P_\bot=0$ with dissipation $q\neq 0$. It should be stressed that in the dissipative case, the pure dust condition, $P_r=P_\bot=0$, does not imply vanishing four-acceleration $a$, as it can be seen from (\ref{29}).

From (\ref{27bis}) it follows at once
\begin{equation}
\ddot R=R^{\prime \prime},
\label{37bis}
\end{equation}
whose general solution is of the form
\begin{equation}
R=c_3\Psi(t+r)+c_4\Phi(t-r),
\label{50bbis}
\end{equation}
where $c_3$ and $c_4$ are constants.
Considering (\ref{37bis}) together with  (\ref{24})  and (\ref{26}) produces
\begin{equation}
2\pi \mu=-\frac{R^{\prime \prime}}{R}=-\frac{\ddot R}{R},
\label{38bis}
\end{equation}
and  (\ref{29}) in our case reduces to,
\begin{equation}
\dot q+2q\frac{\dot R}{R}=0,
\label{39bis}
\end{equation}
implying
\begin{equation}
q=\frac{f_4}{R^2},
\label{40bis}
\end{equation}
where $f_4(r)$ is an arbitrary function of $r$.

We shall now assume that our model satisfies Darmois conditions on $\Sigma^{(i)}$, then from
 (\ref{j3in}) it follows
\begin{equation}
f_4\stackrel{\Sigma^{(i)}}{=}0.
\label{41bis}
\end{equation}
Next, from  (\ref{7}), (\ref{19} and (\ref{25a}) we have
\begin{equation}
\ddot R=-\frac{m}{R^2},
\label{42bis}
\end{equation}
and evaluating it at the  cavity boundary by using (\ref{junction1i}) we obtain
\begin{equation}
R\stackrel{\Sigma^{(i)}}{=}c_5t+c_6
\label{43bis}
\end{equation}
where $c_5$ and $c_6$ are constants. Observe that this is consistent with (\ref{50bbis}).
Also it follows from (\ref{38bis}) that
\begin{equation}
\mu\stackrel{\Sigma^{(i)}}{=}0.
\label{Nmu}
\end{equation}
A further restriction on $f_4$ may be obtained from  (\ref{28n}), which in our case reads
\begin{equation}
\dot \mu+2\mu \frac{\dot R}{R}+q^{\prime}+2q\frac{R^{\prime}}{R}=0.
\label{50bis}
\end{equation}
Evaluating (\ref{50bis}) on the boundary of the cavity and using (\ref{41bis}) and (\ref{Nmu}), we have
\begin{equation}
f_4^{\prime}\stackrel{\Sigma^{(i)}}{=}0.
\label{51bis}
\end{equation}

Once we have assumed that Darmois conditions are satisfied on $\Sigma^{(i)}$, then  it follows that  they are violated on $\Sigma^{(e)}$.
Indeed, if we assume continuity of the mass function across $\Sigma^{(e)}$, then evaluating (\ref{42bis}) on $\Sigma^{(e)}$ and using (\ref{20}) we obtain
\begin{equation}
\ddot R\stackrel{\Sigma^{(e)}}{=}-\frac{M}{R^2},
\label{45bis}
\end{equation}
which can be integrated to obtain
\begin{equation}
\dot R\stackrel{\Sigma^{(e)}}{=}\pm\left(\frac{2M}{R}+c_7\right)^{1/2},
\label{46bis}
\end{equation}
where $c_7$ is a constant.
Combining (\ref{40bis}) and (\ref{50bis}) we obtain
\begin{equation}
\mu R^2=f_5-t f_4^{\prime},
\label{51bbis}
\end{equation}
where $f_5(r)$ is an arbitrary function of $r$.
Then using (\ref{38bis}) and (\ref{45bis}) in (\ref{51bbis}) and evaluating at $\Sigma^{(e)}$ it follows that
\begin{equation}
R \stackrel{\Sigma^{(e)}}{=}\frac{M}{2\pi(f_5-t f_4^{\prime})},
\label{52bbis}
\end{equation}
and by using (\ref{46bis}) becomes
\begin{equation}
R \stackrel{\Sigma^{(e)}}{=}\mbox{constant},
\label{52bbis}
\end{equation}
thereby ruling out  the possibility of the continuity of the mass function  across $\Sigma^{(e)}$. Observe that in this particular case, there should be always a  shell on $\Sigma^{(e)}$.

The only constraints  imposed on functions $f_4$, $\Psi$ and $\Phi$ are, that they must  be  regular and  positive, so that the regularity and positiveness of $\mu$ is assured for all values of $t$ and $r$.

\subsection{Non geodesic models}
So far all presented models within the thin wall approximation have been geodesic, therefore it would be instructive to present a non geodesic model. For that purpose, we shall invoke an ansatz which proved to be very useful for describing dissipative  collapse \cite{Bonnor}.

Thus, let us assume
\begin{equation}
A=A_0(r), \;\; R=R_0(r)g(t), \label{n21}
\end{equation}
where we take $A_0$ and $R_0$ to describe a static anisotropic perfect fluid whose energy density $\mu_0$ and anisotropic pressures $P_{r0}$ and $P_{\perp 0}$ are given by
\begin{eqnarray}
8\pi\mu_0=-2\frac{R_0^{\prime\prime}}{R_0}
-\left(\frac{R_0^{\prime}}{R_0}\right)^2+\frac{1}{R_0^2}, \label{n22}
\end{eqnarray}
\begin{eqnarray}
8\pi P_{r0}=\left(2\frac{A_0^{\prime}}{A_0}+\frac{R_0^{\prime}}{R_0}\right)\frac{R_0^{\prime}}{R_0}-\frac{1}{R_0^2},
\label{n23}
\end{eqnarray}
\begin{eqnarray}
8\pi P_{\perp 0}=\frac{A_0^{\prime\prime}}{A_0}+\frac{R_0^{\prime\prime}}{R_0}+
\frac{A_0^{\prime}}{A_0}\frac{R_0^{\prime}}{R_0}. \label{n24}
\end{eqnarray}
With (\ref{n21}-\ref{n24}) we can rewrite (\ref{24a}-\ref{27a}) like
\begin{eqnarray}
8\pi\mu=\kappa\mu_0+\frac{1}{A_0^2}\left(\frac{\dot g}{g}\right)^2+\frac{1}{R_0^2}\left(\frac{1}{g^2}-1\right),
\label{n25}
\end{eqnarray}
\begin{eqnarray}
8\pi q=\frac{2}{A_0}\left(\frac{R_0^{\prime}}{R_0}-\frac{A_0^{\prime}}{A_0}\right)\frac{\dot g}{g}, \label{n26}
\end{eqnarray}
\begin{widetext}
\begin{eqnarray}
8\pi P_r=\kappa P_{r0}-\frac{1}{A_0^2}\left[2\frac{\ddot g}{g}+\left(\frac{\dot g}{g}\right)^2\right]
-\frac{1}{R_0^2}\left(\frac{1}{g^2}-1\right), \label{n27}
\end{eqnarray}
\end{widetext}
\begin{eqnarray}
8\pi P_{\perp}=8\pi P_{\perp 0}-\frac{1}{A_0^2}\frac{\ddot g}{g}. \label{n28}
\end{eqnarray}

All models of this kind present  shells  in either $\Sigma ^{(i)}$ or $\Sigma ^{(e)}$.
Indeed, evaluating (\ref{n27}) on $\Sigma^{(i)}$ and assuming Darmois conditions there, we obtain
\begin{equation}
2g{\ddot g}+{\dot g}^2-c_0^2(g^2-1)=0, \label{n34bis}
\end{equation}
where
\begin{equation}
c_0\stackrel{\Sigma^{(i)}}{=}\frac{A_0}{R_0}. \label{n34w}
\end{equation}
Then integration of  (\ref{n34bis}) produces
\begin{equation}
{\dot g}^2=c_0^2\left(\frac{g^2}{3}-1\right)+\frac{c_8}{g}, \label{n35bis}
\end{equation}
where $c_8$ is a constant.
Finally, evaluating the mass function  (\ref{18}) on $\Sigma^{(i)}$, and using (\ref{n35bis}) it follows that
\begin{equation}
g(t)=\mbox{constant}.
\label{n39bis}
\end{equation}
implying the necessary  violation of Darmois conditions on  $\Sigma^{(i)}$ .

Choosing a physical meaningful static (``seed'') solution, it is not difficult to choose function $g$ such that standard energy conditions are satisfied. Thus for example if we demand $g^2<1$ and $\frac{\ddot g}{g}>0$, we assure those conditions for  the models.

\section{CONCLUSIONS}
We have studied in detail the consequences emerging from the purely areal evolution condition. It has been shown that such a condition is  particularly suitable for describing the evolution of a fluid distribution endowed with a  cavity surrounding the centre.

All equations governing the dynamics under such condition have been written down  and some models have been presented.

Some of them satisfy Darmois conditions on both hypersurfaces, $\Sigma^{(i)}$  and $\Sigma^{(e)}$), precluding thereby the appearance of shells on either of these hypersurfaces. More simple models result from relaxing Darmois conditions and adopting Israel junction conditions across shells.

One possible  application  of the presented results which comes to our minds, is the modeling of  evolution of  cosmic voids.   Indeed,  the cavity associated to the  purely areal evolution condition might  be considered as a void precursor.

Voids are, roughly speaking,    underdensity regions in the large-scale
matter distribution in the universe  (see  \cite{Zeldovich}-\cite{N} and references therein).

Their relevance stems from the fact that, as stressed in \cite{vn7}, the actual universe has a
spongelike structure, dominated by  voids.  Indeed, observations suggest \cite{vn8} that some 40-50\% of the present
volume of the universe is in voids of a characteristic scale 30 h$^{-1}$ Mpc, where
h is the dimensionless Hubble parameter, $H_{0}$=100 h km sec$^{-1}$ Mpc$^{-1}$.
However voids of very different scales may be found, from minivoids \cite{3} to supervoids \cite{4}.

It should be emphasized that in general  voids are neither empty nor spherical, either in simulations or in deep redshift surveys. However, for simplicity they are usually   described  as vacuum spherical cavities surrounded by a fluid.

However  we are aware of the fact that cold dark matter at scales of the order of tens of Mpc is non-collisional, so that pressure and heat flux terms are negligible. Therefore, excluding the LTB case, it is not likely that our solutions could be used as toy models for cosmic voids.

Possibly our solutions could be applied as toy models of localized  systems such as supernova explosion models. It is worth mentioning that for these scenarios, the Kelvin-Helmholtz phase is of the greatest relevance \cite{b}.

At any rate our purpose here has not been to generate specific models of any observed void, but rather to call the attention to the potential of the purely  areal evolution condition for  such a modeling, providing all necessary equations for their description.

\section*{Acknowledgments.}
We would like to thank an anonymous referee for his very thorough report, which substantially helped us to improve our manuscript.
This research has made use of NASA's Astrophysics Data System.

\end{document}